# MINING THE DATA FROM DISTRIBUTED DATABASE USING AN IMPROVED MINING ALGORITHM

J. Arokia Renjit

Asst. Professor/ CSE Department, Jeppiaar Engineering College,
Chennai, TamilNadu,India – 600119.

Dr.K.L.Shunmuganathan

Professor & Head, Department of CSE, RMK Engineering College,
TamilNadu , India – 601 206.

*Abstract*--**Association rule mining is an active data mining research area and most ARM algorithms cater to a centralized environment. Centralized data mining to discover useful patterns in distributed databases isn't always feasible because merging data sets from different sites incurs huge network communication costs. In this paper, an Improved algorithm based on good performance level for data mining is being proposed. In local sites, it runs the application based on the improved LMatrix algorithm, which is used to calculate local support counts. Local Site also finds a centre site to manage every message exchanged to obtain all globally frequent item sets. It also reduces the time of scan of partition database by using LMatrix which increases the performance of the algorithm. Therefore, the research is to develop a distributed algorithm for geographically distributed data sets that reduces communication costs, superior running efficiency, and stronger scalability than direct application of a sequential algorithm in distributed databases**.

I. INTRODUCTION

Most existing parallel and distributed ARM algorithms are based on a kernel that employs the well-known Apriori algorithm [1]. Directly adapting an Apriori algorithm will not significantly improve performance over frequent item sets generation or overall distributed ARM performance. In distributed mining, synchronization is implicit in message passing, so the goal becomes communication optimization. Data decomposition is very important for distributed memory[2]. Therefore, the main challenge for obtaining good performance on distributed mining is to find a good data decomposition among the nodes for good load balancing, and to minimize communication.

Distributed ARM algorithms aim to generate rules from different data sets spread over various geographical site hence, they require external communications throughout the entire process [3].. They must reduce communication costs so that generating global association rules costs less than combining the participating sites' data sets into a centralized site[4]. Mining association rules is to generate all association rules that have support and confidences are larger than the user-specified minimum support and minimum confidence respectively [5]. The main challenges include work-load balancing, synchronization, communication minimization, finding good data layout, data decomposition, and disk I/O minimization, which is especially important for DARM.

## II. LITERATURE SURVEY

*The Count Distribution (CD) Algorithm*

CD algorithm uses the sequential Apriori algorithm in a parallel environment and assumes datasets are horizontally partitioned among different sites[6]. At each iteration, it generates the candidate sets at every site by applying the Apriori-gen function on the set of frequent itemsets found at the previous iteration. Every site then computes the local support counts of all these candidate sets and broadcasts them to all the other sites. Subsequently, all the sites can find the globally frequent itemsets for that iteration, and then proceed to the next iteration. This algorithm has a simple communication scheme for count exchange. However, it also has the similar problems of higher number of candidate sets and larger amount of communication overhead. It does not use the memory of the system effectively.

*The Fast Distributed Mining Algorithm*

FDM generates fewer candidates than CD, and use effective pruning techniques to minimize the messages for the support exchange step. In each site, FDM finds the local support counts and prunes all infrequent local support counts[7]. After completing local pruning, instead of broadcasting the local counts of all candidates as in CD, they send the local counts to polling site. FDM's main advantage over CD is that it reduces the communication overhead to O ($|Cp|*n$), where $|Cp|$ and n are potentially frequent candidate item sets and the number of sites, respectively[8]. When different sites have nonhomogeneous data sets, the number of disjoint candidate itemsets among them is frequent, and FDM generates fewer candidate itemsets compared to CD

## III. PROPOSED SYSTEM

Mining Association Rules

Efficient algorithms for mining frequent itemsets are crucial for mining association rules as well as for many other data mining tasks. Methods for mining frequent itemsets have been implemented using a prefix-tree structure, known as an FP-tree, for storing compressed information about frequent itemsets. Numerous experimental results have demonstrated that these algorithms perform extremely well. In this paper, we present a novel FP-array technique that greatly reduces the need to traverse FP-trees, thus obtaining significantly improved performance for FP-tree-based algorithms. Our technique works especially well for sparse data sets.Furthermore, we present new algorithms for mining all, maximal, and closed frequent itemsets. The results show that our methods are the fastest for many cases. Even though the algorithms consume much memory when the data sets are sparse, they are still the fastest ones when the minimum support is low.

The L-Matrix Algorithm

Algorithm L-Matrix minimizes the communication overhead. Our solution also reduces the size of average transactions and datasets that leads to reduction of scan time. It minimizes the number of candidate sets and exchange messages by local and global pruning. Reduces the time of scan partition databases to get support counts by using a compressed matrix-L-Matrix, which is very effective in increasing the performance. Finds a centre site to manage every the message exchanges to obtain all globally frequent item sets, only O(n) messages are needed for support count exchange. It has superior running efficiency, lower communication cost and stronger scalability that direct application of a sequential algorithm in distributed databases.

This new algorithm LMatrix is used to achieve maximum efficiency of algorithms..The transaction database is first created to develop the L-Matrix. A LMatrix is an object-by-variable compressed structure. Transaction database is a binary matrix where the rows represent transactions and columns represent alarms. The partitioned databases need to be scanned only once to convert each of them to the local LMatrix. The local LMatrix is read to find support counts instead of scanning the partition databases time after time, which will save a lot of memory. The proposed algorithm can be applied to the mining of association rules in a large centralized database by partitioning the database to the nodes of a distributed system. This is particularly useful if the data set is too large for sequential mining.

LMatrix implementation

The algorithm is implemented with the help of the following supermarket example. Let the supermarket contains five items namely coffee, tea, milk, bread, butter which are represented as A,B,C,D and E respectively and transactions are being done in the following manner. Let us consider three transactions. The first transaction consists of items coffee, tea, and milk. The second transaction consists of items coffee, tea, bread, butter. The third transaction consists of coffee, milk, butter.The LMatrix and the transaction table would look like the one given below.

Then we can obtain the support count of 'A' by accumulating the numbers of '1' in the first column. Then counting the numbers of '1' in Metavector A & C we get the support of AC is 2.

$$\begin{pmatrix} 1 & 1 & 1 & 0 & 0 \\ 1 & 1 & 0 & 1 & 1 \\ 1 & 0 & 1 & 0 & 1 \end{pmatrix}$$

| TRANSACTION ID | ITEM |
|---|---|
| 1 | ABC |
| 2 | ABDE |
| 3 | ACE |

Improved Mining Algorithm

For a site $S_i$, if an itemset X is both locally and globally frequent at site $S_i$, we say that X is heavy at site $S_i$.

A. *Algorithm to compute frequent Itemset in Local Sites.*

1. While flag $_i$ = true, find heavy itemsets at site $s_i$. Then generate the candidate sets using Apriori algorithm.
2. For each candidate set at $s_i$, prune away candidate sets whose max count value is less than s * D, where s is min support and D is partition size of the distributed database.
3. Read LMatrix to compute the local support count of the remaining candidate set. Locally frequent candidate set items are put in $LL_k$
4. Send the candidate sets in $LL_k$ to center sites to collect their global support counts.
5. If $s_i$ receives a count request of itemset X from center site, it reads LMatrix again to obtain support

counts of X and sents it back to centre site else it receives globally frequent itemsets and their support counts.

B. *Algorithm to compute globally frequent Itemset in Central Sites.*

1. Center site receives all $LL_k$ sent to it from the partition sites. When $LL_k = \emptyset$, set flag = false. For every candidate set $X \in LL_k$, it finds the list of originating sites.
2. If all partition sites are in the list of X, put X in $L_k$. Else calculate X.MaxCount and prune away those X whose X.MaxCount < s*D
3. Then broadcast the remaining candidate sets to the other sites not on the list to collect the support counts.
4. Center site receives the local support counts back and adds together and if X.count >= s*D, put it also in $L_k$.
5. Center site then numbers all $X \in L_k$ from 1 to m. X is frequent only when its (k-1) subsets are frequent. If $|L_k| < k+1$, set flag = false.
6. Finally when flag = true, it broadcasts the globally frequent itemsets, together with their global support counts to all the sites and find the heavy itemsets in each site $s_i$.

IV. EXPERIMENTAL RESULTS

If a item is being selected among items A,B,C,D,E in that particular transaction then a count of 1 is incremented for each item. Then a combination of items is being chosen and if it occurs in a particular transaction then a count of 1 is incrementally added to this and the item sets which is less than the minimum support count is removed from the list. After that a combination of three item is chosen then a count of 1 is incremented if it occurs in a particular transaction and items sets having maximum support is the result of the transaction. We get Result== [AC, AE, CE]. In the above result, it is true fact that item D is not in the list of frequent item sets and so it is eliminated and again the above step continues with the help of the items in the list. So the steps above is done locally and now global pruning is done that takes frequent item sets from the both nodes and would result in a final result [A, B, C, E]. So we get the list of items which are locally frequent at site $s_i$ and also globally frequent as follows.

[Coffee, Tea, Milk, Butter]

A   Coffee
B   Tea
C   Milk
E   Butter

The following graphs have been drawn to see the performance of the algorithm in terms of execution time with respect to various minimum supports and database sizes

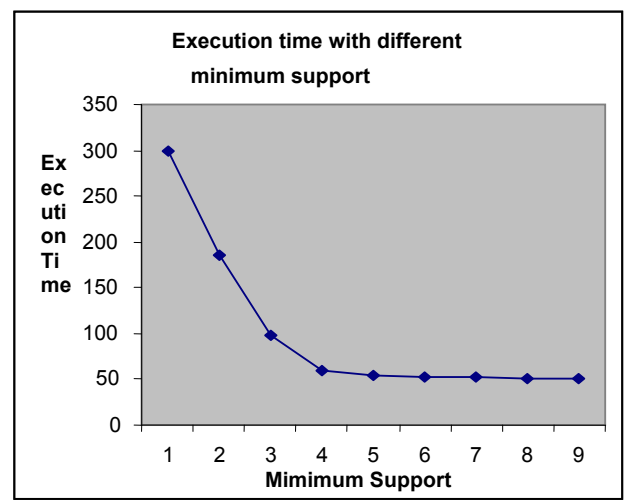

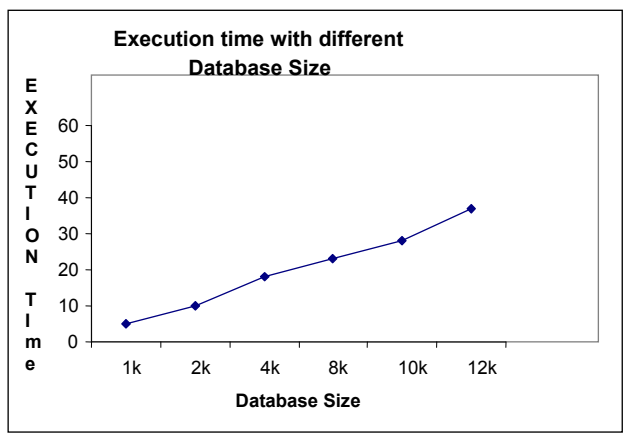

The final transaction table which contains the frequent itemsets alone will look like this.

| TRANS ID | ITEM ID |
|---|---|
| 1 | A |
| 2 | A |
| 3 | A |
| 1 | B |
| 2 | B |
| 1 | C |
| 3 | C |
| 2 | E |
| 3 | E |

V. CONCLUSION

We have developed an efficient algorithm for mining association rules in distributed databases which reduces communication costs and takes away the overhead of combining the partition database sites datasets into a centralized site. It also has the advantage of reduced size of messages passed through the network. It also reduces the time of scan of partition database by using LMatrix which increases the performance of the algorithm. Furthermore, Improved mining algorithm can be applied to the mining of association rules in a large centralized database by partitioning the database to the nodes of a distributed system. This is particularly useful if the data set is too large for sequential mining.

ACKNOWLEDGEMENT

We take immense pleasure in thanking our Chairman Dr. Jeppiaar M.A, B.L, Ph.D, the Directors of Jeppiaar Engineering College Mr. Marie Wilson, B.Tech, MBA.,(Ph.D) Mrs. Regeena Wilson, B.Tech, MBA., (Ph.D) and the Principal Dr. Sushil Lal Das M.Sc(Engg.), Ph.D for their continual support and guidance. We would like to extend our thanks to my guide, our friends and family members without whose inspiration and support our efforts would not have come to true. Above all, we would like to thank God for making all our efforts success.

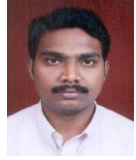

J.ArokiaRenjit B.E.,M.E.,(Ph.D) works as Assistant Professor in Jeppiaar Engineering College and he has more than 8 years of teaching experience. His areas of specializations are Networks, Artificial Intelligence, Software Engineering.

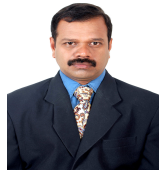

Dr. K.L. Shanmuganathan B.E, M.E.,M.S.,Ph.D works as the Professor & Head of CSE Department of RMK Engineering College, Chennai, TamilNadu, India. He has more than 18 years of teaching experience and his areas of specializations are Artificial Intelligence, Networks and  DBMS